# Quantum ground-mode computation with static gates

**Giuseppe Castagnoli ‡ & David Ritz Finkelstein §**

**Abstract.** We develop a computation model for solving Boolean networks by implementing wires through quantum ground-mode computation and gates through idenities following from angular momentum algebra and statistics. Gates are represented by three-dimensional (triplet) symmetries due to particle indistinguishability and are identically satisfied throughout computation being constants of the motion. The relaxation of the wires yields the network solutions. Such gates cost no computation time, which is comparable with that of an easier Boolean network where all the gate constraints implemented as constants of the motion are removed. This model computation is robust with respect to decoherence and yields a generalized quantum speed-up for all NP problems.

‡ Information Technology Division and Quantum Labs, Elsag spa, 16154 Genova, Italy
§ School of Physics, Georgia Institute of Technology, Atlanta, GA 30332, USA

## 1. Introduction

The prevailing approach to quantum computation is an evolution of classical reversible-algorithmic computation (a sequence of elementary logically reversible transformations, represented by unitary transformations). As well known, entanglement, interference and measurement yield in principle dramatic "speed-ups" over the corresponding classical algorithms in solving some problems. In spite of this outstanding result, it is in general recognized that this form of computation faces two, possibly basic, difficulties. Its speed-ups rely on quantum interference, which requires computation reversibility. Decoherence may then limit computation size below practical interest. Only two speed-ups of practical interest have been found so far (factoring and database search), and none since 1996.

Reversible-algorithmic computation is not the most general form of quantum computation. Its limitations justify reconsidering quantum ground-mode computation (Castagnoli 1998, Farhi et al. 2001, Kadowaki 2002, among others), a formerly neglected approach, still believed to be mathematically intractable. It is a quantum version of classical ground-state computation (Kirkpatrick & Selman 1994, among others), which is a well-developed approach competitive with algorithmic computation for solving Boolean networks (a most general problem).

A Boolean network is a set of nodes (Boolean variables) variously connected by gates and wires that impose equations on the variables they connect (fig. 1). A Boolean assignment satisfying all gates and wires is a network solution.

In quantum ground-mode computation, one sets up a quantum network whose energy is minimum when all gates and wires are satisfied. Coupling the network with a heat-bath of suitably decreasing temperature, relaxes the network to its ground mode, a mixture of solutions (we assume there is at least one). Measuring the node variables (Hermitian operators with eigenvalues 0 and 1) yields a solution. It is believed that this form of computation yields a (still ill-defined) speed-up over its classical counterpart: quantum tunneling reduces network trapping in local energy minima (e.g. Kadowaki 2002). However, long simulation times seriously limit research on this approach.

Here we develop a hybrid mode of computation. Only wires are implemented by ground-mode computation. Gates are always identically satisfied throughout the computation being constants of the motion. Gates are implemented by quantum symmetries due to particle indistinguishability. We show that relaxation-computation time is comparable with that an easier (loosely constrained) logical network where all the gate constraints implemented by quantum symmetries are removed. We plausibly conjecture that for this computation mode all hard to solve (NP) networks become easy (P). Decoherence is not a problem in principle since the network mode can be a mixture.

This model computation still belongs to the realm of principles, like other literature on quantum ground-mode computation, while algorithmic-reversible computation is now almost a technology. Nevertheless, it is worth starting over with a new approach that might overcome fundamental limitations of algorithmic computation.

## 2. Computation model

We use a network normal form with just wires and triodes (fig.1). Each triode $\tau$ – properly a partial gate – connects three nodes labeled $\tau, x$ - $\tau, y$ - $\tau, z$ (replaced by collective indices in fig.1) imposing the equation $q_{\tau,x} + q_{\tau,y} + q_{\tau,z} = 2$, where q's are Boolean variables and + denotes arithmetical sum. The three solutions are the rows of table I. Each wire $w$ connects two nodes $i, j$ imposing $q_i = q_j$, table II. The network exemplified in fig. 1, with $Q = 6$ nodes, $W = 4$ wires, and $T = 2$ triodes, has just one solution: $q_3 = q_5 = 0$, $q_1 = q_2 = q_4 = q_6 = 1$.

| $q_{\tau,x}$ | $q_{\tau,y}$ | $q_{\tau,z}$ |
|---|---|---|
| 0 | 1 | 1 |
| 1 | 0 | 1 |
| 1 | 1 | 0 |

Table I

| $q_i$ | $q_j$ |
|---|---|
| 0 | 0 |
| 1 | 1 |

Table II

| $q_{\tau,x}$ | $q_{\tau,y}$ | $q_{\tau,z}$ | sy |
|---|---|---|---|
| 0 | 0 | 0 | s |
| 0 | 1 | 1 | t |
| 1 | 0 | 1 | t |
| 1 | 1 | 0 | t |

Table III

Fig. 1. A network

All network nodes belong to triodes. The node variables are physically represented by associating each triode $\tau$ with a proton pair with spin vectors $\frac{1}{2}\sigma_{\tau,1}, \frac{1}{2}\sigma_{\tau,2}$ in units $\hbar = 1$. Two independent spin vectors have total spin $\frac{1}{2}(\sigma_{\tau,1} + \sigma_{\tau,2})$ with $s_{\tau,z} = \pm 1, 0$, and define three commuting Hermitian operators with eigenvalues 0 and 1: $q_{\tau,x} = s^2_{\tau,x}$, $q_{\tau,y} = s^2_{\tau,y}$, $q_{\tau,z} = s^2_{\tau,z}$, each corresponding to a network node. For the composition of angular momentum, the eigenvalues of the three operators of each proton pair must satisfy the XOR gate equation (table III). Its four rows correspond to the singlet and the three triplet pair modes of $\tau$, spanning the Hilbert space $\mathcal{H}^{(4)}_\tau$. We use $\mathcal{H}^{(4)}_N = \bigotimes_{\tau=1}^T \mathcal{H}^{(4)}_\tau$ as the network space.

To model the triodes physically we assume that the spatial wave function of each proton pair $\tau$ is frozen antisymmetric in a stable ground mode (like in orthohydrogen nuclei) throughout the computation. Angular momentum composition and triplet symmetry are extra-dynamical: the triode Hamiltonians are zero in $\mathcal{H}^{(4)}_N$. Let $\mathcal{H}^{(3)}_\tau$ be the space spanned by the three triplet modes of triode $\tau$. $\mathcal{H}^{(3)}_N = \bigotimes_{\tau=1}^T \mathcal{H}^{(3)}_\tau \subset \mathcal{H}^{(4)}_N$ is the network subspace with all triodes satisfied.

The wire frustration Hamiltonian in $\mathcal{H}^{(4)}_N$ is $H^{(4)}_N = g \sum_{\{i,j\}} (q_i - q_j)^2$, where $\{i,j\}$ is the set of all wires and $g$ a coefficient to provide the dimension of energy. $H^{(4)}_N$ is quadrilinear in the $s_{\tau,w}$. By using two auxiliary spin 1/2 variables for each wire, the wire Hamiltonian can be made bilinear, as in the Ising model (Castagnoli & Finkelstein 2002, briefly "I"). $H^{(4)}_N$ is symmetric under all $X_{12,\tau}$, the exchange operators of the two protons of each triode $\tau$, since the $q$'s are. Triplet symmetry is thus a constant of motion of $H_N^{(4)}$.

The heat-bath is a photon filled cavity, with Hilbert space $\mathcal{H}_B$. $\mathcal{H}^{(4)} := \mathcal{H}_N^{(4)} \otimes \mathcal{H}_B$ is the "system" (=network+bath) space, $\mathcal{H}^{(3)} := \mathcal{H}_N^{(3)} \otimes \mathcal{H}_B$ its subspace with triplet symmetry (all triodes satisfied). We denote by $H_B^{(4)}(t)$ the heat-bath Hamiltonian and define the network-bath coupling in $\mathcal{H}^{(4)}$ by $H_I^{(4)}(t) = g \sum_\tau \left[ \vec{B}_\tau(t) \cdot \vec{\sigma}_{\tau,1} + \vec{B}_\tau(t) \cdot \vec{\sigma}_{\tau,2} \right]$. Each proton spin is coupled to a small random Gaussian time-varying magnetic field $\vec{B}_\tau(t)$ of the photon at the site of the spin. Indistinguishability requires that the two protons of the same triode $\tau$ experience the same magnetic field (we assume no overlap between spatial wave functions of different proton pairs). Therefore triplet symmetry (triodes satisfaction) is a constant of motion of $H_I^{(4)}(t)$, and also of the system Hamiltonian $H^{(4)}(t) = H_N^{(4)} + H_B^{(4)}(t) + H_I^{(4)}(t)$ ($H_N^{(4)}$ is already symmetric). Let $|\psi, t\rangle$ be the mode of the system, developed by $H^{(4)}(t)$. The relaxation of the network mode is described by the statistical operator $\rho_N(t) := \text{Tr}_B(|\psi, t\rangle \langle \psi, t|)$, where $\text{Tr}_B$ means trace over the bath degrees of freedom. If $\rho_N(t)$ starts in $\mathcal{H}^{(3)}$, under $H^{(4)}(t)$ it remains in it.

With a suitable time-variation of $\vec{B}_\tau(t)$, $H^{(4)}(t)$ relaxes the network to its zero point. A direct estimate of this relaxation time is likely mathematically intractable, and simulation is very lengthy. We take a shortcut that also sheds light on the nature of this hybrid computation. We compare the network relaxation time with that of an easier network obtained by replacing all triodes (Table I) by XOR gates (Table III): as if proton indistinguishability were suspended – each proton pair replaced by a deuteron. Restriction to $\mathcal{H}^{(3)}$ vanishes: a network of XOR gate and wires is loosely constrained and easy to solve. In particular all $q_i = 0$ is always a solution. A XOR network is solvable in poly($Q$) time in classical computation and, reasonably, also in the present form of computation (for more detail, see I). The asymmetric Hamiltonian of the comparison system in $\mathcal{H}^{(4)}$ is $H^\blacklozenge(t) = H_N^{(4)} + H_B^{(4)}(t) + H_I^\blacklozenge(t)$, where $H_I^\blacklozenge(t) = g \sum_\tau \left[ \vec{B}_{\tau,1}(t) \cdot \vec{\sigma}_{\tau,1} + \vec{B}_{\tau,2}(t) \cdot \vec{\sigma}_{\tau,2} \right]$ is the asymmetric coupling. Now we have two independent random Gaussian time-varying magnetic fields at each proton site, such that $\vec{B}_\tau(t) = \left[ \vec{B}_{\tau,1}(t) + \vec{B}_{\tau,2}(t) \right]/2$ is the actual heat-bath (the sum of two Gaussian distributions is Gaussian). Let $|\varphi, t\rangle$ be the mode of the comparison system, developed by $H^\blacklozenge(t)$. The comparison network relaxation is described by $\rho_N^\blacklozenge(t) := \text{Tr}_B(|\varphi, t\rangle \langle \varphi, t|)$. Let $P := 2^{-T} \prod_{\tau=1}^T [1 + X_{12,\tau}]$ be the network symmetrization operator. It projects $\mathcal{H}^{(4)}$ on $\mathcal{H}^{(3)}$. Clearly $PH_I^\blacklozenge(t) P = H_I^{(4)}(t)$, thus $PH^\blacklozenge(t) P = H^{(4)}(t)$ ($H_N^{(4)}$ is already symmetric, $P$ is the identity in $\mathcal{H}_B$).

We show that the development of the actual system (hard triode network and bath) is the continuous projection on $\mathcal{H}^{(3)}$ of the development of the comparison system (easy XOR network and bath) in $\mathcal{H}^{(4)}$. Let $|\psi, t\rangle \subset \mathcal{H}^{(3)}$ so that $P |\psi, t\rangle = |\psi, t\rangle$. Under $H^{(4)}(t)$, it develops into $|\psi, t + dt\rangle = \left(1 - iH^{(4)}(t) dt\right) |\psi, t\rangle$; under $H^\blacklozenge(t)$ into $|\varphi, t + dt\rangle := \left(1 - iH^\blacklozenge(t) dt\right) |\psi, t\rangle$, in general non-symmetric. We restore particle indistinguishability by projecting $|\varphi, t + dt\rangle$ on $\mathcal{H}^{(3)}$, symmetrizing it: $P |\varphi, t + dt\rangle = \left(P^2 - iPH^\blacklozenge(t) P\right) |\psi, t\rangle = \left(1 - iH^{(4)}(t) dt\right) |\psi, t\rangle = |\psi, t + dt\rangle$. Thus the continuous projection of the comparison development yields the actual development.

Computation time is by assumption poly($Q$) for the comparison easy network. To estimate that of the actual hard network, we decompose a $\Delta T$ into $N = \Delta T / \Delta t$ consecutive time slices $\Delta t_i \equiv [t_i, t_{i+1}]$ of equal length $\Delta t$. Within each $\Delta t_i$, we consider the relaxation $\rho_N^\blacklozenge(t)$ of the comparison XOR network in $\mathcal{H}^{(4)}$. At the end of each $\Delta t_i$, we project $\rho_N^\blacklozenge(t)$ on $\mathcal{H}^{(3)}$, then take the limit $\Delta t \to 0$. This yields the relaxation of the actual network $\rho_N(t)$.

Within each $\Delta t_i$ we consider the decomposition $\rho_N^\blacklozenge(t) := \rho_0(t) + \rho_F(t) + \rho_V(t)$. $\rho_0(t)$ describes networks with satisfied triodes and wires, namely solutions of the actual network; its probability is $p_0(t) := \mathrm{Tr}\rho_0(t)$. $\rho_F(t)$ describes networks with satisfied triodes and at least one frustrated wire; $p_F(t) := \mathrm{Tr}\rho_F(t)$. $\rho_V(t)$ describes networks with at least one violated triode, wires are either satisfied or frustrated; $p_V(t) := \mathrm{Tr}\rho_V(t)$. All possible modes of the comparison network are considered, thus $p_0(t) + p_F(t) + p_V(t) = 1$. $p_V(t)$ goes to zero with $\Delta t$ and is annihilated by each projection.

The actual network-bath interaction soon randomly generates a $\rho_0(t_h)$, a mixture of solutions of the actual network, with extremely small probability $p_0(t_h) = O\left(1/2^Q\right)$. For a given confidence level, $t_h$ does not depend on $Q$. For $t > t_h$ we apply the projection method. $p_0(t_h) = O\left(1/2^Q\right)$ becomes the nucleus of condensation of the network solutions.

Within each and every $\Delta t_i$, we take a constant-average logarithmic rate of decrease $k$ of the comparison network frustration energy: $E_N(t_{i+1}) = (1 - k\Delta t) E_N(t_i)$. There is no error in taking a constant rate – see later. The relaxation time constant $1/k$ is by assumption poly($Q$). Since $E_N(t) = Tr\rho_F(t) H_N^{(4)}$ (no contribution from $\rho_0(t)$, second order infinitesimal from $\rho_V(t)$) $E_N(t)$ and $p_F(t)$ go to zero together (see I for more detail). Thus on average:

$$p_F(t_{i+1}) = (1 - k\Delta t) p_F(t_i). \tag{1}$$

$p_F(t)$ decrease implies an increase of $p_0(t) + p_V(t)$. It is reasonable and conservative to consider the increase of $p_V(t)$ dominant. The relaxation of the comparison network is quicker because triodes can be violated. Note that we compare relaxation rates, not directions: the comparison network can head toward $\mathcal{H}_N^{(4)} \sim \mathcal{H}_N^{(3)}$, the actual network remains in $\mathcal{H}_N^{(3)}$. Furthermore, $H_N^{(4)}$ does not couple $\rho_0(t)$ with $\rho_F(t)$ or $\rho_V(t)$ ($H_N^{(4)}\rho_0(t) = \rho_0(t)H_N^{(4)} = 0$). Therefore $p_0(t)$ neither decreases nor increases on average. Since $p_F(t)$ decreases and $p_0(t)$ does not, the ratio $p_F(t)/p_0(t)$ decreases. When we project on $\mathcal{H}_N^{(3)}$ at the end of $\Delta t_i$, we remain with a smaller $p_F(t)$ and a larger $p_0(t)$ (probability of solutions of the actual network). For (1), with $p_0(t)$ unaltered within each $\Delta t_i$ and not too close to 1 (say $p_0(t) < 3/10$, so that $p_F(t)$ remains close to 1), and $\Delta t \to 0$, we have approximately for the actual network:

$$p_0(t_h + \Delta T) \approx p_0(t_h) e^{k\Delta T} \approx \frac{1}{2^Q} e^{k\Delta T}, \tag{2}$$

as readily checked (see also I). The probability of having solutions of the actual network becomes $O(1)$ in a $\Delta T \approx Q/k = Q$poly($Q$)=poly($Q$).

Using a different $k_i$ for each $\Delta t_i$, with average $k$ ($\sum_i k_i \Delta t_i = k\Delta T$), yields the same result: $e^{k\Delta T}$ in (2) should be replaced by $\prod_i e^{k_i \Delta t_i} = e^{k\Delta T}$.

In conclusion, particle indistinguishability yields a new form of quantum computation where the gates of a Boolean network are always satisfied as constants of the motion. This model computation appears to be promising for achieving: (i) robust quantum computation and (ii) a generalized speed-up in all NP problems.

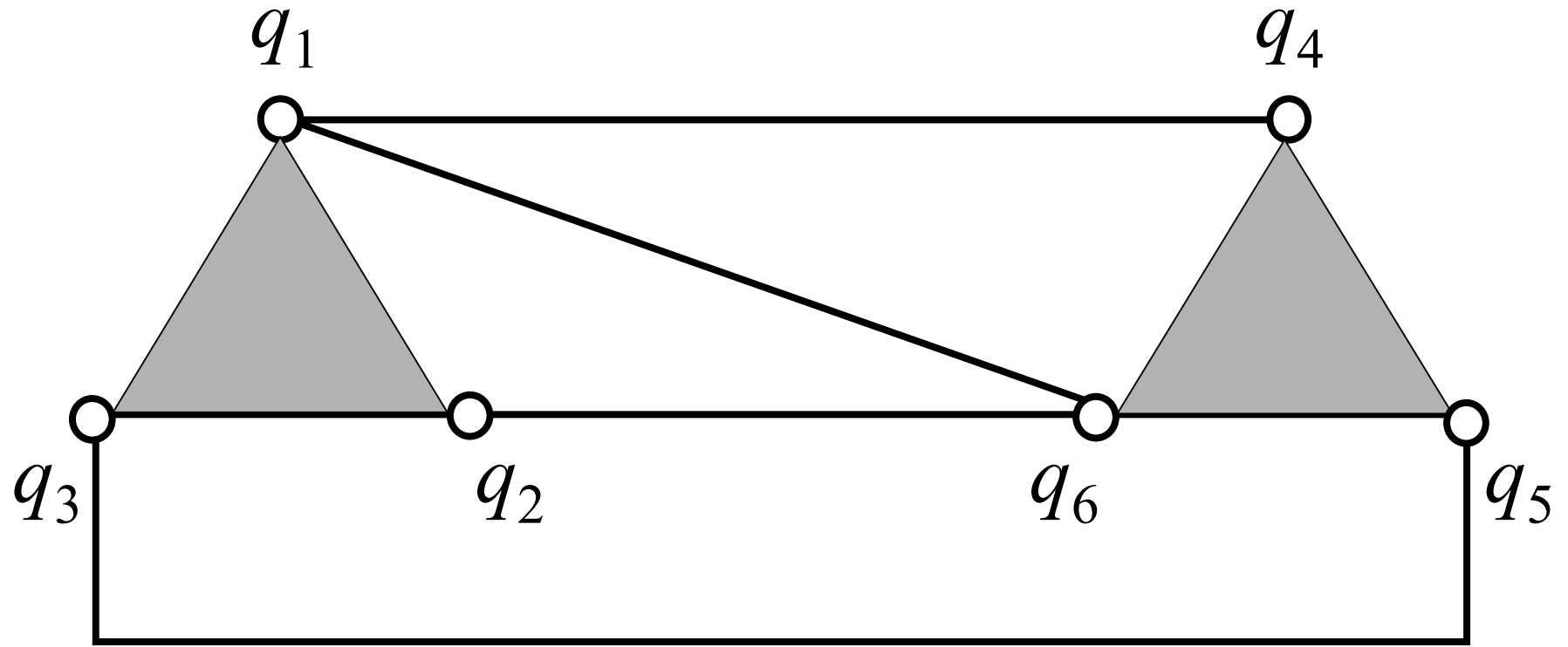

Fig. 1. A network